\documentclass[a4paper, 12pt, oneside]{article}
\usepackage[cp1251]{inputenc}
\usepackage[english, russian]{babel}
\usepackage{ graphics,amsfonts, amsmath,bm,amssymb, array, eucal}
\usepackage[dvips]{graphicx}
\usepackage[flushleft,small]{caption2}

\makeatletter
\newcommand{\const}{\mathop{\rm const\, }}
\renewcommand{\Re}{\mathop{\rm Re\,}}
\renewcommand{\Im}{\mathop{\rm Im\,}}
\makeatother
\renewcommand{\baselinestretch}{1.2}
\begin{document}
\newcommand{\mc}[1]{\mathcal{#1}}
\newcommand{\E}{\mc{E}}
\thispagestyle{empty}
\large

\renewcommand{\abstractname}{Abstract }
\renewcommand{\refname}{\begin{center} REFERENCES\end{center}}

 \begin{center}
\bf Longitudinal dielectric permeability into quantum non-degenerate
and maxwellian plasma with frequency of collisions proportional to the
module of a wave vector
\end{center}\medskip
\begin{center}
  \bf A. V. Latyshev\footnote{$avlatyshev@mail.ru$} and
  A. A. Yushkanov\footnote{$yushkanov@inbox.ru$}
\end{center}\medskip

\begin{center}
{\it Faculty of Physics and Mathematics,\\ Moscow State Regional
University, 105005,\\ Moscow, Radio str., 10--A}
\end{center}\medskip

\begin{abstract}
Formulas for the longitudinal dielectric permeability in quantum
non-degenerate and maxwellian collisional plasma with the frequency of
collisions proportional to the module of the wave
vector, in approach Мермина, are received.
Equation of Shr\"{o}dinger---Boltzmann with integral
of collisions relaxation type in Mermin's appro\-ach is applied.

It is spent numerical and graphic
comparison of the real and imaginary parts of dielectric function
of non-degenerate and maxwellian collisional quantum plasma
with a constant and a variable frequencies of collisions.
It is shown, that the longitu\-dinal
dielectric function weakly depends on a wave vector.


\end{abstract}

\begin{center}
{\bf 1. Introduction}
\end{center}

In  Klimontovich and Silin's work  \cite{Klim} expression
for longitudinal and transverse dielectric permeability of quantum
collisionless plasmas has been received.

Then in Lindhard's work \cite{Lin} expressions
has been received  also for the same characteristics of quantum
collisionless plasma.

By Kliewer and Fuchs \cite{Kliewer} it has been shown, that
direct generalisation of formulas of Lindhard  on a case of collisionless
plasmas, is incorrectly.
This lack for the longitudinal dielectric
permeability has been eliminated in work of Mermin \cite{Mermin} for
collisional plasmas.
In this work of Mermin \cite{Mermin} on the basis of the analysis
of a nonequilibrium matrix
density in $ \tau $-approach expression for
longitudinal dielectric permeability of quantum collisional plasmas
in case of constant frequency of collisions of particles of plasma
has been announced.

For collisional plasmas correct formulas longitudinal and transverse
electric conductivity and dielectric permeability are received
accordingly in works \cite{Long} and \cite{Trans}. In these works
kinetic  Wigner---Vlasov---Boltzmann equation
in relaxation approximation in coordinate space was used.

In work \cite{Trans2} the formula for the transverse electric
conductivity of quantum collisional plasmas with use of the kinetic
Shr\"{o}dinger---Boltzmann equation in Mermin's approach  (in space of
momentum) has been deduced.

In work \cite{Long2} the formula for the longitudinal dielectric
permeability of quantum collisional plasmas with use of the kinetic
Shr\"{o}dinger---Boltzmann equation in approach of Mermin (in space of
momentum) with any variable frequency of collisions depending from
wave vector  has been deduced.

In the present work on the basis of results from our previous work
\cite{Long2} formulas for longitudinal dielectric permeability
in quantum collisional plasma with frequency of collisions,
proportional to the module of a wave vector are received.
The modelling is thus used Shr\"{o}dinger---Boltzmann equation
in relaxation approximation.

In our work \cite{Lat2007} formulas for longitudinal and transverse
electric conductivity in the classical collisional
gaseous (maxwellian) plasma with frequency of collisions
of plasma particles proportional to the
module particles velocity  have been deduced.

Research of
skin-effect in classical collisional gas plasma with frequency
of collisions proportional to the module particles velocity
has been carried out in work \cite{Lat2006}.

Let's notice, that interest to research of the phenomena
in quantum plasma grows in last years \cite{Manf} -- \cite{Ropke}.

\begin{center}
\bf 1. Longitudinal dielectric function of quantum collisional
plasma with variable collisional frequency
\end{center}

In work \cite{Long} longitudinal dielectric function of the quantum
collisional plasmas with frequency of collisions,
proportional to the module of a wave vector has been received
$$
\varepsilon_l({\bf q},\omega,\nu)=1+\dfrac{4\pi e^2}{q^2}\Big[B({\bf q},\omega+
i \bar \nu)+ \hspace{6cm}
$$
$$+ib_{\bar \nu}({\bf q},\omega+i \bar \nu)
\dfrac{b({\bf q},0)-b({\bf q},\omega+i\bar \nu)}
{\omega b({\bf q},0)+ib_{\omega,\bar \nu}({\bf q},\omega+i\bar \nu)}\Big].
\eqno{(1.1)}
$$\medskip

In the formula (1.1) $e $ is the electron charge, $ {\bf q} $ is the wave
vector, $ \omega $ is the frequency of oscillations of an electromagnetic field,
$ \nu ({\bf k}) $ is the frequency of collisions of particles of
plasma,

$$
\bar\nu=\bar \nu({\bf k,q})=\bar \nu({\bf k}+\dfrac{{\bf q}}{2},
{\bf k}-\dfrac{{\bf q}}{2})=\dfrac{\nu\big({\bf k}+\dfrac{{\bf q}}{2}\big)+
\nu\big({\bf k}-\dfrac{{\bf q}}{2}\big)}{2},
\eqno{(1.2)}
$$\medskip

$$
B({\bf q},\omega+i\bar\nu)=\int\dfrac{d^3k}{4\pi^3}\Big(f_{{\bf k+q}/2}-
f_{{\bf k-q}/2}\Big)\Xi(\omega+i\bar \nu({\bf k+q}/2,{\bf k-q}/2)),
\eqno{(1.3)}
$$\medskip
$$
b({\bf q},\omega+i\bar\nu)=\int\dfrac{d^3k}{4\pi^3}\Big(f_{{\bf k+q}/2}-
f_{{\bf k-q}/2}\Big)\Xi(\omega+i\bar \nu({\bf k+q}/2,{\bf k-q}/2))\times
$$\medskip
$$
\times\dfrac{\bar \nu({\bf k+q}/2,{\bf k-q}/2)}{\omega+i
\bar \nu({\bf k+q}/2,{\bf k-q}/2)},
\eqno{(1.4)}
$$\medskip
$$
b({\bf q},0)=\int\dfrac{d^3k}{4\pi^3}\Big(f_{{\bf k+q}/2}-
f_{{\bf k-q}/2}\Big)\Xi(0)\dfrac{\bar \nu({\bf k+q}/2,{\bf k-q}/2)}{\omega+i
\bar \nu({\bf k+q}/2,{\bf k-q}/2)},
\eqno{(1.5)}
$$\medskip
$$
b_{\bar \nu}({\bf q},\omega+i\bar\nu)=
\int\dfrac{d^3k}{4\pi^3}\Big(f_{{\bf k+q}/2}-
f_{{\bf k-q}/2}\Big)\Xi(\omega+i\bar \nu({\bf k+q}/2,{\bf k-q}/2))\times
$$\medskip
$$
\times{\bar \nu({\bf k+q}/2,{\bf k-q}/2)},
\eqno{(1.6)}
$$\medskip
$$
b_{\omega,\bar \nu}({\bf q},\omega+i\bar\nu)=
\int\dfrac{d^3k}{4\pi^3}\Big(f_{{\bf k+q}/2}-
f_{{\bf k-q}/2}\Big)\Xi(\omega+i\bar \nu({\bf k+q}/2,{\bf k-q}/2))\times
$$\medskip
$$
\times\dfrac{\bar \nu^2({\bf k+q}/2,{\bf k-q}/2)}{\omega+i
\bar \nu({\bf k+q}/2,{\bf k-q}/2)},
\eqno{(1.7)}
$$\medskip

In integrals (1.3) -- (1.7) the following designations are accepted
$$
\Xi(\omega+i\bar \nu({\bf k+q}/2,{\bf k-q}/2))=$$$$=
\dfrac{1}{\E_{{\bf k-q}/2}-\E_{{\bf k+q}/2}+\hbar[\omega+i \bar
 \nu({\bf k+q}/2,{\bf k-q}/2)]},
$$\medskip
$$
f_{{\bf k}}=\dfrac{1}{1+\exp\Big(\dfrac{\E_{{\bf k}}-\mu}{k_BT}\Big)},
$$\medskip
$$
\E_{{\bf k\pm q}/2}=\dfrac{\hbar^2}{2m}\Big({\bf k}\pm\dfrac{{\bf
q}}{2}\Big)^2.
$$

Here $m $ is the electron mass, $k_B $ is the Boltzmann constant,
$ \mu $ is the chemical potential of molecules of gas, $ \hbar $ ie the
Planck's constant.

Let's show, that at $ \nu ({\bf k}) = \nu =\const $, i.e. at a constant
collisional frequency the formula (1.1) passes in the known Mermin's formula
\cite{Mermin}

$$
\varepsilon_l^{\rm Mermin}=1+\dfrac{4\pi e^2}{q^2}\dfrac{(\omega+i \nu)
B({\bf q},\omega+i \nu)B({\bf q},0)}{\omega B({\bf q},0)+i \nu
B({\bf q},\omega+i \nu)}.
\eqno{(1.8)}
$$

In (1.8) the following designations are used

$$
B({\bf q},\omega+i \nu)=\int \dfrac{d^3k}{4\pi^3}
(f_{{\bf k+q}/2}-f_{{\bf k-q}/2})\Xi(\omega+ i \nu),
\eqno{(1.9)}
$$
$$
B({\bf q},0)=\int \dfrac{d^3k}{4\pi^3}
(f_{{\bf k+q}/2}-f_{{\bf k-q}/2})\Xi(0),\qquad
$$
$$
\Xi(\omega+i \nu)=\dfrac{1}
{\E_{{\bf k-q}/2}-\E_{{\bf k+q}/2}+\hbar(\omega+i \nu)}.
$$

Let's notice, that at $ \nu({\bf k}) \equiv \nu $, $ \bar\nu({\bf k, q}) \equiv
\nu $, and we receive following equalities
$$
B({\bf q},\omega+i\bar \nu)\equiv B({\bf q},\omega+i \nu),
$$
$$
b({\bf q},\omega+i\bar \nu)=\dfrac{\nu}{\omega+i \nu}B({\bf q},\omega+i \nu),
$$
$$
b({\bf q},0)=\dfrac{\nu}{\omega+i \nu}B({\bf q},0),
$$
$$
b_{\bar \nu}({\bf q},\omega+i\bar \nu)=\nu B({\bf q},\omega+i \nu),
$$
$$
b_{\omega,\bar \nu}({\bf q},\omega+i\bar \nu)=\dfrac{\nu^2}{\omega+i \nu}
B({\bf q},\omega+i \nu).
$$

It is as a result received, that
$$
\varepsilon_l({\bf q},\omega,\nu)=1+\dfrac{4\pi e^2}{q^2}
B({\bf q},\omega+i \nu)\Big[1+
$$
$$
+i \nu\dfrac{B({\bf q},0)-
B({\bf q},\omega+i \nu)}{\omega B({\bf q},0)+
i \nu B({\bf q},\omega+i \nu)}\Big] \equiv \varepsilon_l^{\rm Mermin}
({\bf q},\omega,\nu).
$$\medskip

Each of integrals (1.3) -- (1.7) we will break into a difference
of two integrals. In each of two integrals it is realizable the obvious
linear replacement of variables. It is as a result received, that
$$
B({\bf q},\omega+i\bar \nu)=\int \dfrac{d^3k}{4\pi^3}f_{{\bf k}}
\Big[\Xi(\omega+i\bar\nu({\bf k,k-q}))-\Xi(\omega+i\bar\nu({\bf k+q,k}))\Big],
\eqno{(1.10)}
$$

$$
b({\bf q},\omega+i\bar \nu)=\int \dfrac{d^3k}{4\pi^3}f_{{\bf k}}
\Big[\dfrac{\bar\nu({\bf k,k-q})}{\omega+i\bar\nu({\bf k,k-q})}
\Xi(\omega+i\bar\nu({\bf k,k-q}))-$$$$-\dfrac{\bar\nu({\bf k+q,k})}
{\omega+i\bar\nu({\bf k+q,k}))}
\Xi(\omega+i\bar\nu({\bf k+q,k}))\Big],
\eqno{(1.11)}
$$ \medskip

$$
b({\bf q},0)=\int \dfrac{d^3k}{4\pi^3}f_{{\bf k}}
\Big[\dfrac{\bar\nu({\bf k,k-q})}{\omega+i\bar\nu({\bf k,k-q})(\E_{{\bf k-q}}-
\E_{{\bf k}})}
-$$$$-\dfrac{\bar\nu({\bf k+q,k})}
{\omega+i\bar\nu({\bf k+q,k}))(\E_{\bf k}-\E_{{\bf k+q}})}\Big],
\eqno{(1.12)}
$$

$$
b_{\bar \nu}({\bf q},\omega+i\bar \nu)=\int \dfrac{d^3k}{4\pi^3}f_{{\bf k}}
\Big[{\bar\nu({\bf k,k-q})}
\Xi(\omega+i\bar\nu({\bf k,k-q}))-$$$$-{\bar\nu({\bf k+q,k})}
\Xi(\omega+i\bar\nu({\bf k+q,k}))\Big],
\eqno{(1.13)}
$$

$$
b_{\omega,\bar\nu}({\bf q},\omega+i\bar \nu)=\int \dfrac{d^3k}{4\pi^3}f_{{\bf k}}
\Big[\dfrac{\bar\nu^2({\bf k,k-q})}{\omega+i\bar\nu({\bf k,k-q})}
\Xi(\omega+i\bar\nu({\bf k,k-q}))-$$$$-\dfrac{\bar\nu^2({\bf k+q,k})}
{\omega+i\bar\nu({\bf k+q,k})}
\Xi(\omega+i\bar\nu({\bf k+q,k}))\Big].
\eqno{(1.14)}
$$ \medskip

In integrals (1.10) -- (1.14) following designations are accepted
$$
\bar \nu({\bf k,k-q})=\dfrac{\nu({\bf k})+\nu({\bf k-q})}{2},
$$
$$
\bar \nu({\bf k+q,k})=\dfrac{\nu({\bf k+q})+\nu({\bf k})}{2},
$$
$$
\Xi(\omega+i\bar\nu({\bf k,k-q})=\dfrac{1}{\E_{\bf k-q}-\E_{{\bf k}}+
\hbar[\omega+i\bar\nu({\bf k,k-q})]},
$$
$$
\Xi(\omega+i\bar\nu({\bf k+q,k})=\dfrac{1}{\E_{\bf k}-\E_{{\bf k+q}}+
\hbar[\omega+i\bar\nu({\bf k+q,k})]}.
$$ \medskip

\begin{center}
  \bf 2. Longitudinal dielectric function of the quantum
  collisional non-degenerate plasmas with frequency of collisions,
  proportional to the module of a wave vector
\end{center}

Let's consider the frequency of collisions proportional
to the momentum module, or, that all the same, to the module of a wave vector:
$$
\nu({\bf k})=\nu_0|{\bf k}|.
$$

Then
$$
\bar\nu({\bf k}_1,{\bf k}_2)=\dfrac{\nu({\bf k}_1)+\nu({\bf k}_2)}{2}=
\dfrac{\nu_0}{2}\Big(|{\bf k}_1|+|{\bf k}_2|\Big)
$$
and
$$
\bar \nu({\bf k,q})=\bar \nu\Big({\bf k}+\dfrac{{\bf q}}{2},
{\bf k}-\dfrac{{\bf q}}{2} \Big)=\dfrac{\nu_0}{2}\Big(\Big|{\bf k}+
\dfrac{{\bf q}}{2}\Big|+\Big|{\bf k}-\dfrac{{\bf q}}{2}\Big|\Big).
$$

The quantity $ \nu_0$ we take in the form $ \nu_0 =\dfrac {\nu} {k_T} $, where
$k_T $ is the thermal wave number, $k_T =\dfrac {mv_T} {\hbar} $,
$ \hbar $ is the Planck's constant, $v_T $ is the thermal electron velocity. Now
$$
\nu({\bf k})=\dfrac{\nu}{k_T}|{\bf k}|.
\eqno{(2.1)}
$$

Let's notice, that at $k=k_T $: $ \nu (k_T) = \nu $.
So, further in formulas (1.1) -- (1.7) frequency
of collisions according to (2.1) is equal:
$$
\bar\nu({\bf k,q})=\dfrac{\nu}{2k_T}\Big(\Big|{\bf k}+
\dfrac{{\bf q}}{2}\Big|+\Big|{\bf k}-\dfrac{{\bf q}}{2}\Big|\Big).
\eqno{(2.2)}
$$

Instead of a vector $ {\bf k} $ we will enter the new dimensionless wave
vector of integration
$$
{\bf K}=\dfrac{{\bf k}}{k_T},\qquad d^3k=k_T^3\,d^3K.
$$

Let's enter also a new wave vector
$$
{\bf Q}=\dfrac{{\bf q}}{k_T}.
$$

At the specified replacement of variables we have
$$
f_{{\bf k}}=\dfrac{1}{1+e^{{\bf K}^2-\alpha}}=f_{\rm {\bf K}}.
$$

According to the specified replacement of variables further it is received
$$
\bar \nu({\bf k,k-q})=\dfrac{\nu}{2}\Big(|{\bf K}|+|{\bf K-Q}|\Big),
$$
$$
\bar \nu({\bf k+q,k})=\dfrac{\nu}{2}\Big(|{\bf K+Q}|+|{\bf K}|\Big),
$$

$$
\E_{{\bf k-q}}-\E_{{\bf k}}+\hbar[\omega+i\bar\nu({\bf k,k-q})]=
$$
$$
=\dfrac{\hbar^2}{2m}\Big[({\bf k-q})-{\bf k}^2\Big]+
\hbar[\omega+i\bar\nu({\bf k,k-q})]=
$$
$$
=-2\E_{\rm T}Q\Big(K_x-\dfrac{Q}{2}\Big)+\hbar[\omega+i\bar\nu({\bf k,k-q})]=
$$
$$
=-2\E_{\rm T}Q\Big(K_x-\dfrac{Q}{2}-\dfrac{z^-}{Q}\Big).
$$

Here
$$
{\bf Q}=Q(1,0,0),\qquad z^-=x+iy\rho^-,\qquad
x=\dfrac{\omega}{k_{\rm T}v_{\rm T}},\qquad
y=\dfrac{\nu}{k_{\rm T}v_{\rm T}},
$$

$$
\rho^-=\dfrac{1}{2}\Big(|{\bf K}|+|{\bf K-Q}|\Big)=
$$
$$
=\dfrac{1}{2}\Big[\sqrt{K_x^2+K_y^2+K_z^2}+
\sqrt{(K_x-Q)^2+K_y^2+K_z^2}\,\Big].
$$

Similarly we receive, that
$$
\E_{{\bf k}}-\E_{{\bf k+q}}+\hbar[\omega+i\bar\nu({\bf k,k-q})]=
$$

$$
=-2\E_{\rm T}Q\Big(K_x+\dfrac{Q}{2}-\dfrac{z^+}{Q}\Big),\qquad
z^+=x+iy\rho^+,
$$

$$
\rho^+=\dfrac{1}{2}\Big(|{\bf K}|+|{\bf K+Q}|\Big)=
$$
$$
=\dfrac{1}{2}\Big[\sqrt{K_x^2+K_y^2+K_z^2}+
\sqrt{(K_x+Q)^2+K_y^2+K_z^2}\,\Big].
$$

Let's pass to new variables in integrals (1.10) -- (1.14).
We receive following equalities. For integral (1.10) it is had
$$
B({\bf q},\omega+i\bar\nu)=-\dfrac{k_T^3}{8\pi^3\E_TQ}B(Q,z^{\pm}),
$$
where
$$
B(Q,z^{\pm})=\int f_{{\bf K}}\Big[\dfrac{1}{K_x-Q/2-z^-/Q}-\dfrac{1}
{K_x+Q/2-z^+/Q}\Big]d^3K.
$$

For integral (1.11) it is received
$$
b({\bf q},\omega+i\bar\nu)=-\dfrac{yk_T^3}{8\pi^3\E_TQ}b(Q,z^{\pm}),
$$
where
$$
b(Q,z^{\pm})=\int f_{{\bf K}}\Big[\dfrac{\rho^-}{z^-(K_x-Q/2-z^-/Q)}-
\dfrac{\rho^+}{z^+(K_x+Q/2-z^+/Q)}\Big]d^3K.
$$

For integral (1.12) it is received
$$
b({\bf q},0)=-\dfrac{yk_T^3}{8\pi^3\E_TQ}b(Q,0^{\pm}),
$$
where
$$
b(Q,0^{\pm})=\int f_{{\bf K}}\Big[\dfrac{\rho^-}{z^-(K_x-Q/2)}-
\dfrac{\rho^+}{z^+(K_x+Q/2)}\Big]d^3K.
$$
For integral (1.13) it is received
$$
b_{\bar\nu}({\bf q},\omega+i\bar\nu)=-\dfrac{yk_T^4v_T}{8\pi^3\E_{\rm T}Q}
b_{\bar\nu}(Q,z^{\pm}),
$$
where
$$
b_{\bar\nu}(Q,z^{\pm})=\int f_{{\bf K}}\Big[\dfrac{\rho^-}{K_x-Q/2-z^-/Q}-
\dfrac{\rho^+}{K_x+Q/2-z^+/Q}\Big]d^3K.
$$

At last, for integral (1.14) it is similarly received
$$
b_{\omega,\bar\nu}({\bf q},\omega+i\bar\nu)=-\dfrac{y^2k_T^4v_T}{8\pi^3\E_TQ}
b_{\omega,\bar\nu}(Q,z^{\pm}),
$$
where
$$
b_{\omega,\bar \nu}(Q,z^{\pm})=
\int f_{{\bf K}}\Big[\dfrac{{\rho^-}^2}{z^-(K_x-Q/2-z^-/Q)}-
\dfrac{{\rho^+}^2}{z^+(K_x+Q/2-z^+/Q)}\Big]d^3K.
$$

Let's substitute the received equalities in the formula (1.1). We receive
the expression for longitudinal dielectric function
$$
\varepsilon_l(Q,x,y)=1-\dfrac{3x_p^2}{4\pi Q^3}\Big[B(Q,z^{\pm})+\hspace{6cm}
$$$$+
iyb_{\bar\nu}(Q,z^{\pm})\dfrac{b(Q,0^{\pm})-b(Q,z^{\pm})}{xb(Q,0^{\pm})+
iyb_{\omega,\bar\nu}(Q,z^{\pm})}\Big].
\eqno{(2.3)}
$$

Here $x_p$ is the dimensionless plasma (Langmuir) frequency,
$$
x_p=\dfrac{\omega_p}{k_{\rm  T}v_{\rm T}}, \qquad
\omega_p^2=\dfrac{4\pi^2 eN}{m},
$$
$\omega_p$ is the dimension plasma (Langmuir) frequency.

Let's notice, that in case of constant frequency of electron collisions
the quantity $ \rho^{\pm} $ passes in unit. Then
$$
B(Q,z^{\pm})=QB(Q,z), \qquad b(Q,0)=\dfrac{Q}{z}B(Q,0),
$$
$$
b_{\bar\nu}(Q,z^{\pm})=Q B(Q,z),\qquad b_{\omega,\bar\nu}(Q,z^{\pm})=
\dfrac{Q}{z}B(Q,z),
$$
where
$$
B(Q,z)=\int \dfrac{f_{{\bf K}}d^3K}{(K_x-z/Q)^2-(Q/2)^2}.
$$

Substituting these equalities in (2.3), we receive expression
of dielectric function for quantum non-degenerate
collisional plasmas with constant frequ\-ency of collisions
$$
\varepsilon_l(Q,x,y)=1-\dfrac{3x_p^2}{4\pi Q^2}B(Q,z)\Big[1+iy
\dfrac{B(Q,0)-B(Q,z)}{xB(Q,0)+iyB(Q,z)}\Big].
$$

Let's result the formula (2.3) in the calculation form. For this purpose
in the plane $ (K_y, K_z) $ we will pass to polar coordinates
$$
K_y^2+K_z^2=r^2,\qquad dK_ydK_z=rdrd\varphi.
$$

Then
$$
\varepsilon_l(Q,x,y)=1-\hspace{4cm}
$$ \medskip
$$-\dfrac{3x_p^2}{2Q^3}\Big[D(Q,z^{\pm})+
iyd_{\bar\nu}(Q,z^{\pm})\dfrac{d(Q,0)-d(Q,z^{\pm})}{xd(Q,0)+
iyd_{\omega,\bar\nu}(Q,z^{\pm})}\Big].
\eqno{(2.4)}
$$

Here
$$
D(Q,z^{\pm})=\int\limits_{-\infty}^{\infty}dK_x\int\limits_{0}^{\infty}
\Big(\dfrac{1}{K_x-Q/2-z^-/Q}- \hspace{4cm}
$$
$$
\hspace{3cm}-\dfrac{1}{K_x+Q/2-z^+/Q}\Big)
f_{\rm F}(K_x,r,\alpha)rdr,
$$\medskip
$$
z^-=x+iy\rho^-, \qquad
\rho^-=\dfrac{1}{2}\Big(\sqrt{(K_x-Q)^2+r^2}+\sqrt{K_x^2+r^2}\Big),
$$\medskip
$$
z^+=x+iy\rho^+, \qquad
\rho^+=\dfrac{1}{2}\Big(\sqrt{(K_x+Q)^2+r^2}+\sqrt{K_x^2+r^2}\Big),
$$\medskip
$$
f_{\rm
F}(K_x,r,\alpha)=\dfrac{1}{1+\exp\Big(K_x^2+r^2-\alpha\Big)}.
$$

Besides,
$$
d(Q,z^{\pm})=\int\limits_{-\infty}^{\infty}dK_x\int\limits_{0}^{\infty}
\Big(\dfrac{\rho^-}{z^-(K_x-Q/2-z^-/Q)}-\hspace{4cm}
$$
$$-
\dfrac{\rho^+}{z^+(K_x+Q/2-z^+/Q)}\Big)f_{\rm F}(K_x,r,\alpha)rdr,
$$

$$
d(Q,0)=
\int\limits_{-\infty}^{\infty}dK_x\int\limits_{0}^{\infty}
\Big[\dfrac{\rho^-}{(x+iy\rho^-)(K_x-Q/2)}-\hspace{4cm}
$$
$$-
\dfrac{\rho^+}{(x+iy\rho^+)(K_x+Q/2)}\Big]f_{\rm F}(K_x,r,\alpha)rdr,
$$

$$
d_{\bar\nu}(Q,z^{\pm})=\int\limits_{-\infty}^{\infty}dK_x
\int\limits_{0}^{\infty}
\Big(\dfrac{\rho^-}{K_x-Q/2-z^-/Q}-\hspace{4cm}
$$
$$-
\dfrac{\rho^+}{K_x+Q/2-z^+/Q}\Big)f_{\rm F}(K_x,r,\alpha)rdr,
$$
and, at last,
$$
d_{\omega,\bar\nu}(Q,z^{\pm})=
\int\limits_{-\infty}^{\infty}dK_x\int\limits_{0}^{\infty}
\Big(\dfrac{{\rho^-}^2}{z^-(K_x-Q/2-z^-/Q)}-\hspace{4cm}
$$
$$-
\dfrac{{\rho^+}^2}{z^+(K_x+Q/2-z^+/Q)}\Big)f_{\rm F}(K_x,r,\alpha)rdr.
$$

For comparison we take the formula for longitudinal dielektric
func\-ti\-ons in case of constant frequency of collisions of plasma
particles\\
$\nu({\bf k})=\nu=\const$:
$$
\varepsilon_l(Q,x,y)=1-\dfrac{3x_p^2}{4\pi Q^2}\dfrac{(x+iy)B(Q,z)B(Q,0)}
{xB(Q,0)+iyB(Q,z)}.
$$
Here
$$
B(Q,z)=\int\dfrac{f_{{\bf K}}d^3K}{(K_x-z/Q)^2-(Q/2)^2}=
\pi \int\limits_{-\infty}^{\infty}\dfrac{\ln(1+e^{\alpha-K_x^2})dK_x}
{(K_x-z/Q)^2-(Q/2)^2},
$$
$$
B(Q,0)=\pi \int\limits_{-\infty}^{\infty}\dfrac{\ln(1+e^{\alpha-K_x^2})dK_x}
{K_x^2-(Q/2)^2},\qquad z=x+iy.
$$
\medskip

\begin{center}
  \bf 3. Quantum maxwellian collisinal plasmas
\end{center}

In case of quantum maxwellian plasmas $f_{{\bf K}} =e^{-{\bf
K}^2} $. Hence, longitudinal dielectric function
it is calculated again under the formula (2.4) in which now are accepted
following designations
$$
D(Q,z^{\pm})=\int\limits_{-\infty}^{\infty}e^{-K_x^2}dK_x\int\limits_{0}^{\infty}
\Big(\dfrac{1}{K_x-Q/2-z^-/Q}-
$$
$$
-\dfrac{1}{K_x+Q/2-z^+/Q}\Big)
e^{-r^2}rdr,
$$\medskip
$$
z^-=x+iy\rho^-, \qquad
\rho^-=\dfrac{1}{2}\Big(\sqrt{(K_x-Q)^2+r^2}+\sqrt{K_x^2+r^2}\Big),
$$
$$
z^+=x+iy\rho^+, \qquad
\rho^+=\dfrac{1}{2}\Big(\sqrt{(K_x+Q)^2+r^2}+\sqrt{K_x^2+r^2}\Big).
$$

Besides,
$$
d(Q,z^{\pm})=\int\limits_{-\infty}^{\infty}e^{-K_x^2}dK_x
\int\limits_{0}^{\infty}
\Big(\dfrac{\rho^-}{z^-(K_x-Q/2-z^-/Q)}-\hspace{4cm}
$$
$$-
\dfrac{\rho^+}{z^+(K_x+Q/2-z^+/Q)}\Big)e^{-r^2}rdr,
$$

$$
d(Q,0)=
\int\limits_{-\infty}^{\infty}e^{-K_x^2}dK_x\int\limits_{0}^{\infty}
\Big[\dfrac{\rho^-}{(x+iy\rho^-)(K_x-Q/2)}-\hspace{4cm}
$$
$$-
\dfrac{\rho^+}{(x+iy\rho^+)(K_x+Q/2)}\Big]e^{-r^2}rdr,
$$

$$
d_{\bar\nu}(Q,z^{\pm})=\int\limits_{-\infty}^{\infty}e^{-K_x^2}dK_x
\int\limits_{0}^{\infty}
\Big(\dfrac{\rho^-}{K_x-Q/2-z^-/Q}-\hspace{4cm}
$$
$$-
\dfrac{\rho^+}{K_x+Q/2-z^+/Q}\Big)f_{\rm F}(K_x,r,\alpha)e^{-r^2}rdr,
$$
and, finally,
$$
d_{\omega,\bar\nu}(Q,z^{\pm})=
\int\limits_{-\infty}^{\infty}e^{-K_x^2}dK_x\int\limits_{0}^{\infty}
\Big(\dfrac{{\rho^-}^2}{z^-(K_x-Q/2-z^-/Q)}-\hspace{4cm}
$$
$$-
\dfrac{{\rho^+}^2}{z^+(K_x+Q/2-z^+/Q)}\Big)e^{-r^2}rdr.
$$

For comparison we take the formula for the longitudinal dielectric
function in case of constant frequency of collisions of plasma particles\\
 $\nu({\bf k})=\nu
=\const$:
$$
\varepsilon_l(Q,x,y)=1-\dfrac{3x_p^2}{4\pi Q^2}\dfrac{(x+iy)B(Q,z)B(Q,0)}
{xB(Q,0)+iyB(Q,z)}.
$$
Here
$$
B(Q,z)=\int\dfrac{f_{{\bf K}}d^3K}{(K_x-z/Q)^2-(Q/2)^2}=
\pi \int\limits_{-\infty}^{\infty}\dfrac{e^{-K_x^2}dK_x}
{(K_x-z/Q)^2-(Q/2)^2},
$$
$$
B(Q,0)=\pi \int\limits_{-\infty}^{\infty}\dfrac{e^{-K_x^2}dK_x}
{K_x^2-(Q/2)^2},\qquad z=x+iy.
$$

On Figs. 1-4 comparison of the real and imaginary parts
of dielectric function depending on quantity of the dimensionless
wave vector $Q $ (Figs. 1,2) and depending on the dimensionless
quantity of frequency of an electromagnetic field $x $ (Figs. 3,4) is carry out.
Thus
curves 1 correspond to values of frequency
the collisions, proportional to the module of a wave vector; curves
2 correspond to constant frequency of collisions of particles of plasma.
Both curves are constructed at $y=0.01$.
All graphics answer to non-degenerate quantum plasma.
The case of maxwellian  plasmas is considered on Figs. 5-8.

Everywhere more low $x_p=1$, and value of chemical potential equally:
$ \alpha=0$.

On Figs. 9-12 the case of maxwellian plasmas is considered.
On Figs. 9 and 10 comparison of the real parts is considered
at $x=1, 0\le Q\le 3$ (fig. 9)
and imaginary parts of dielectric function at $Q=1, 0\le x\le 3$ (fig. 10)
of quantum collisinal plasmas
with the frequency of collisions proportional to the module of a wave vector.
Curves 1,2,3 answer according to values $y=0.1,0.05,
0.001$.

On Figs. 11 and 12 comparison is carry out of a relative deviation
of real (curves 1) and imaginary parts (curves 2)
of dielectric function from the present work (with frequency of collisions,
proportional to the module of a wave vector) with the corresponding
dielectric Mermin's function (with constant collision
frequency) at the same parametres, and quantity
$y =\dfrac{\nu}{k_Fv_F} =0.01$ is the same. Curves 1 on Figs. 11 and 12
are defined by function
$$
O_r(Q,x,y)=\dfrac{\Re \varepsilon_l^{\rm Mermin}(Q,x,y)-\Re\varepsilon_l(Q,x,y)}
{\Re \varepsilon_l^{\rm Mermin}(Q,x,y)},
$$
and curves  2 are defined by function
$$
O_i(Q,x,y)=\dfrac{\Im \varepsilon_l^{\rm Mermin}(Q,x,y)-\Im\varepsilon_l(Q,x,y)}
{\Im \varepsilon_l^{\rm Mermin}(Q,x,y)}.
$$

\begin{center}
\bf 5. Conclusions
\end{center}

In the present work formulas for the longitudinal
dielectric permeability (dielectric function) in the quantum
collisional non-degenerate plasma (with any degree of degeneration)
and maxwellian plasma are deduced.
Frequency of collisions plasma particles
it is supposed proportional to the module of a wave vector (or an momentum
plasma particles). Graphic research of behaviour
the real and imaginary parts of the found dielectric
functions is carried out.
Comparison of the real and imaginary parts is spent also
the found dielectric function with the corresponding
characteristics of dielectric function with constant collision
frequency.

\begin{figure}[t]\center
\includegraphics[width=16.0cm, height=10cm]{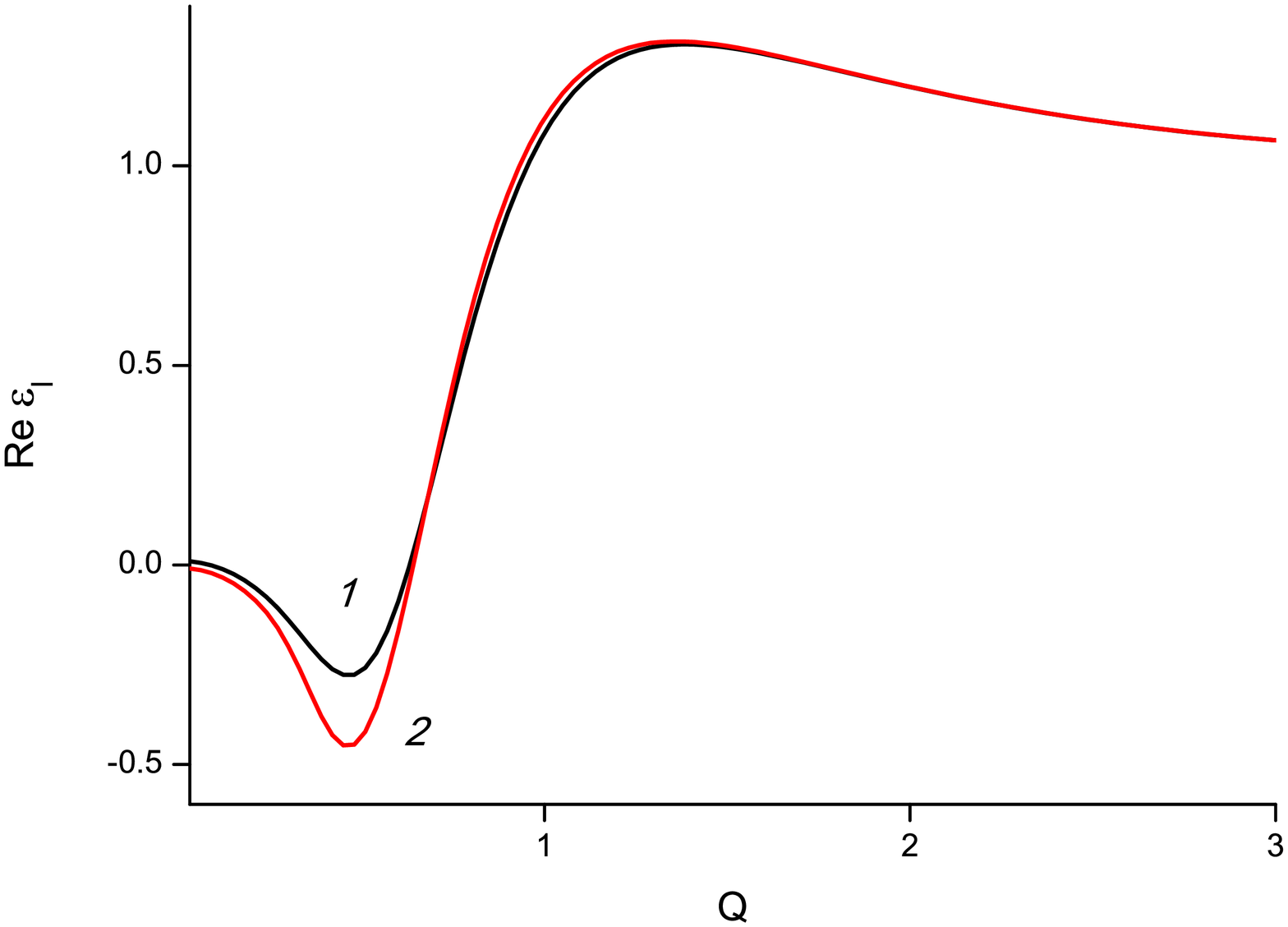}
\center{Fig. 1. Real part of dielectric function,
$x=1, y=0.01$.}
\includegraphics[width=17.0cm, height=10cm]{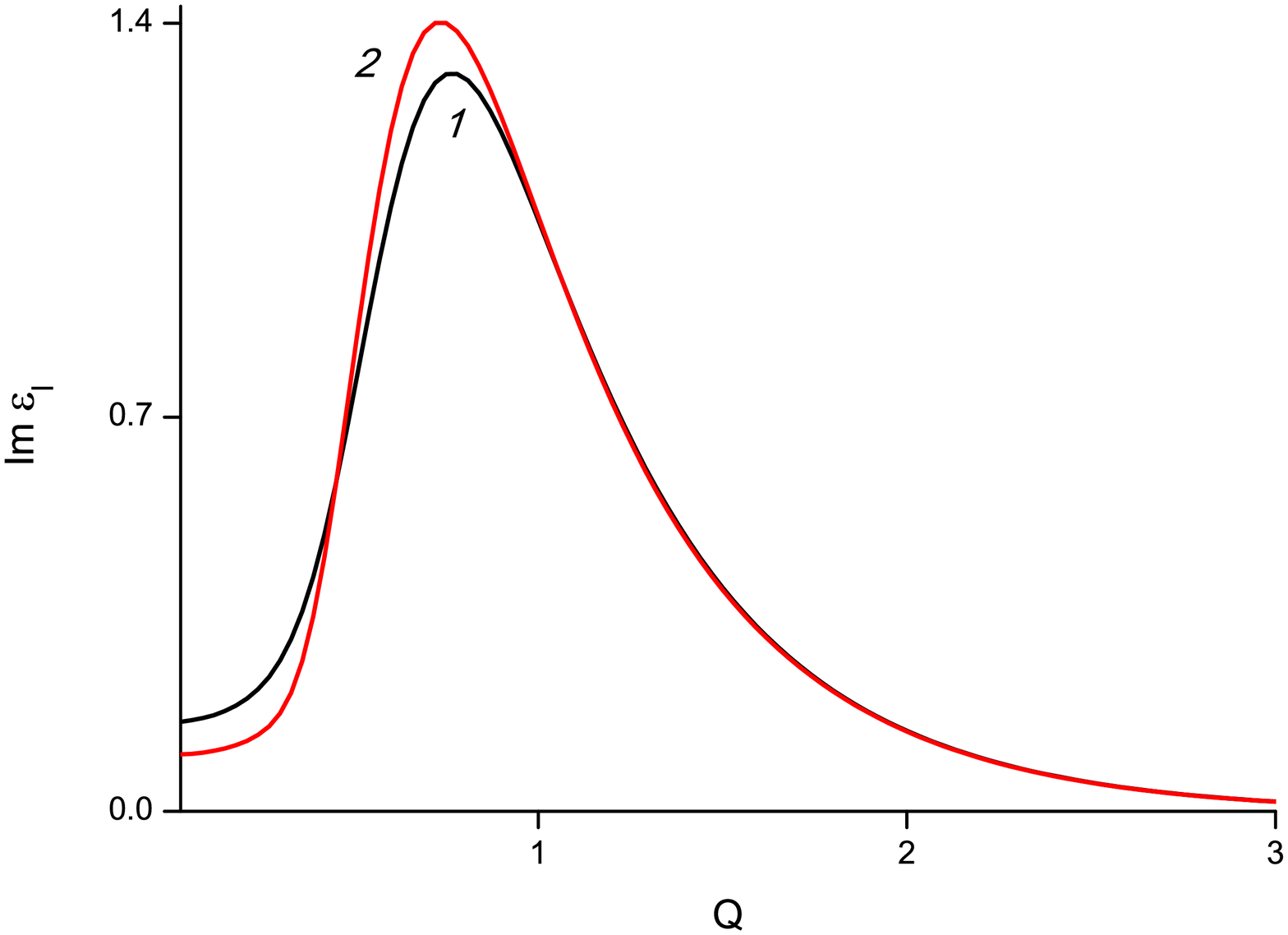}
\center{Fig. 2. Imaginary part of dielectric function,
$x=1, y=0.01$.}
\end{figure}

\begin{figure}[h]\center
\includegraphics[width=16.0cm, height=10cm]{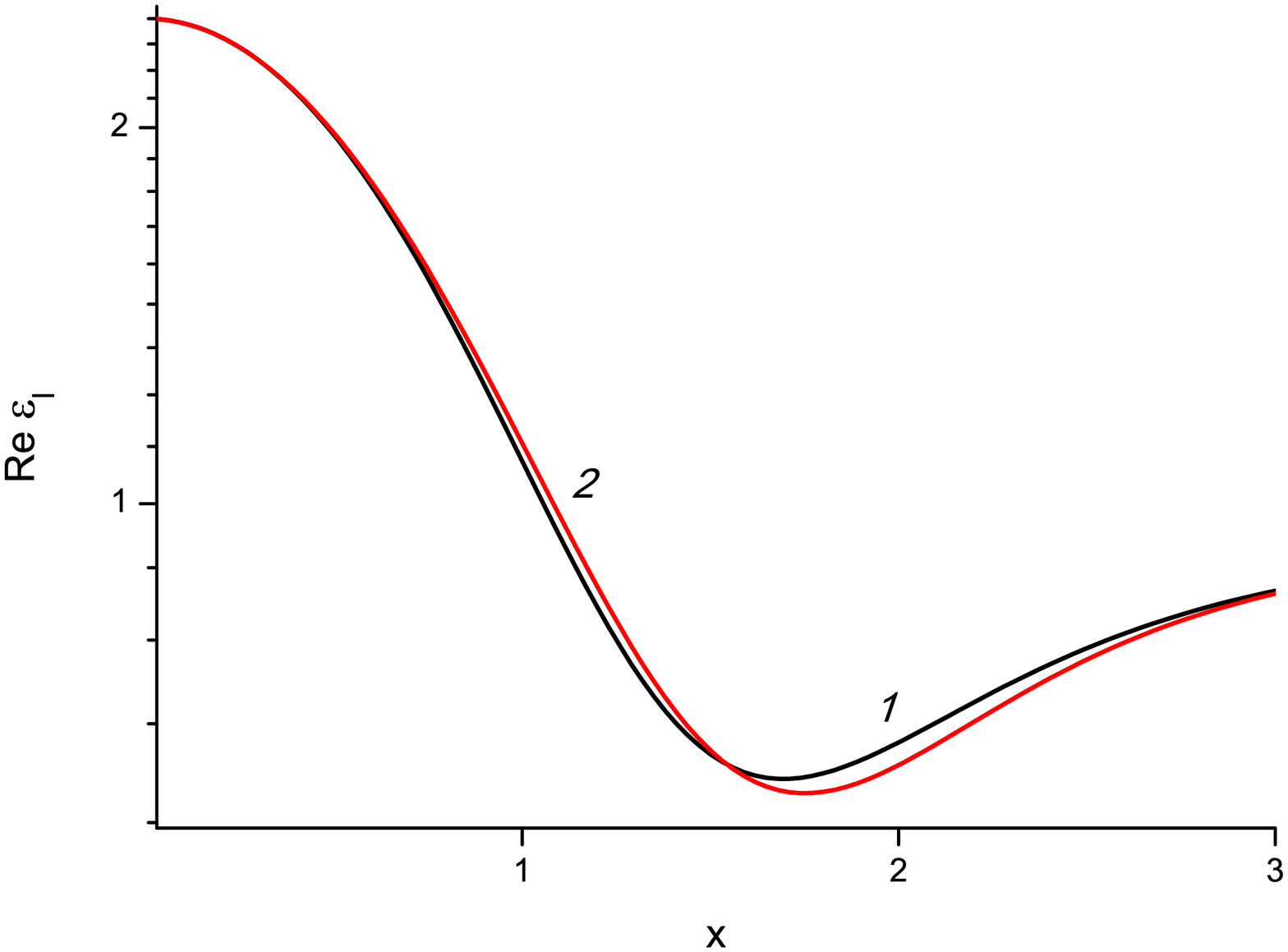}
\center{Fig. 3. Real part of dielectric function,
$Q=1, y=0.01$.}
\includegraphics[width=17.0cm, height=10cm]{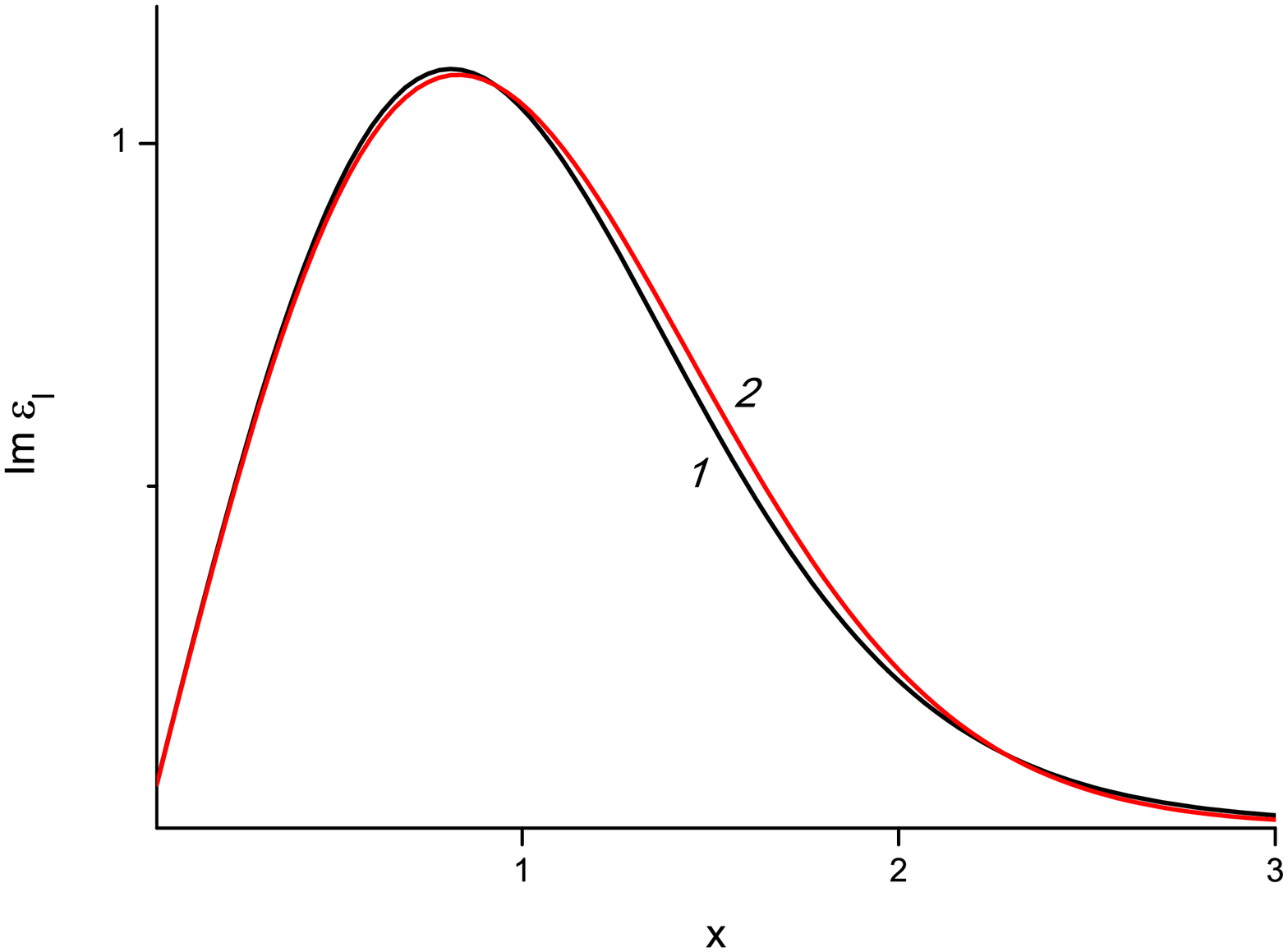}
\center{Fig. 4. Imaginary part of dielectric function,
$Q=1, y=0.01$.}
\end{figure}

\begin{figure}[t]\center
\includegraphics[width=16.0cm, height=10cm]{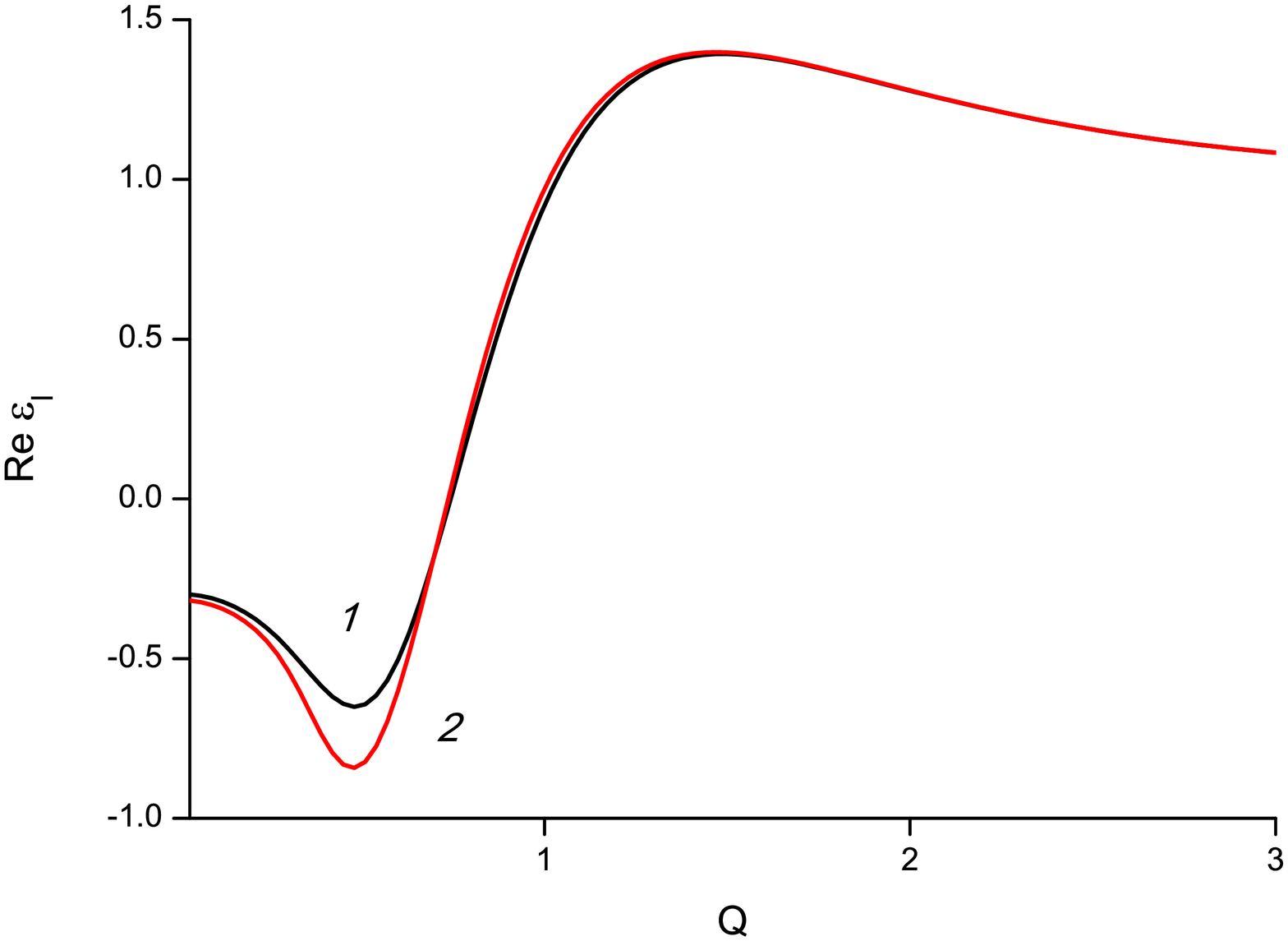}
\center{Fig. 5. Real part of dielectric function,
$x=1, y=0.01$.}
\includegraphics[width=17.0cm, height=10cm]{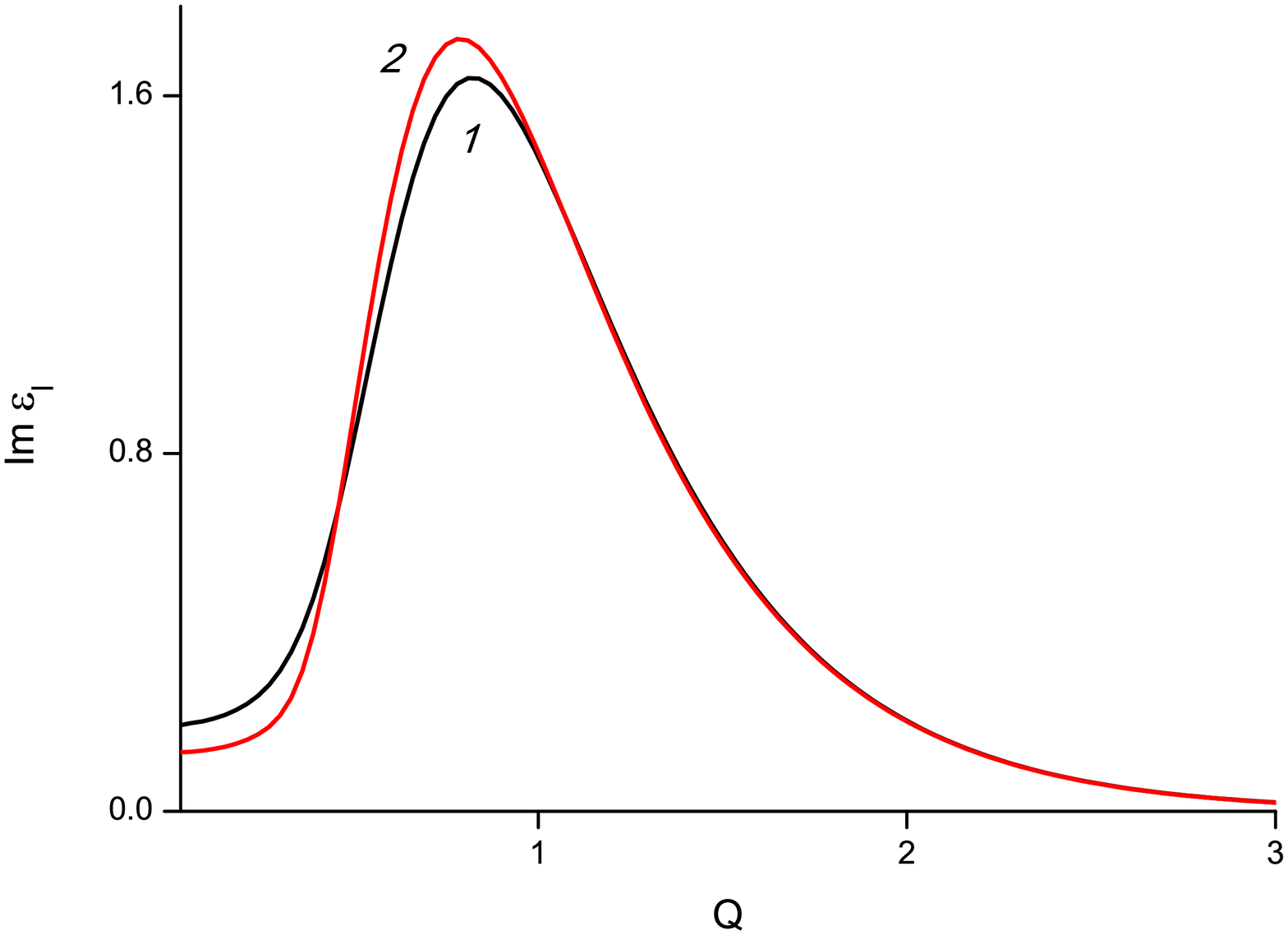}
\center{Fig. 6. Imaginary part of dielectric function,
$x=1, y=0.01$.}
\end{figure}

\begin{figure}[h]\center
\includegraphics[width=16.0cm, height=10cm]{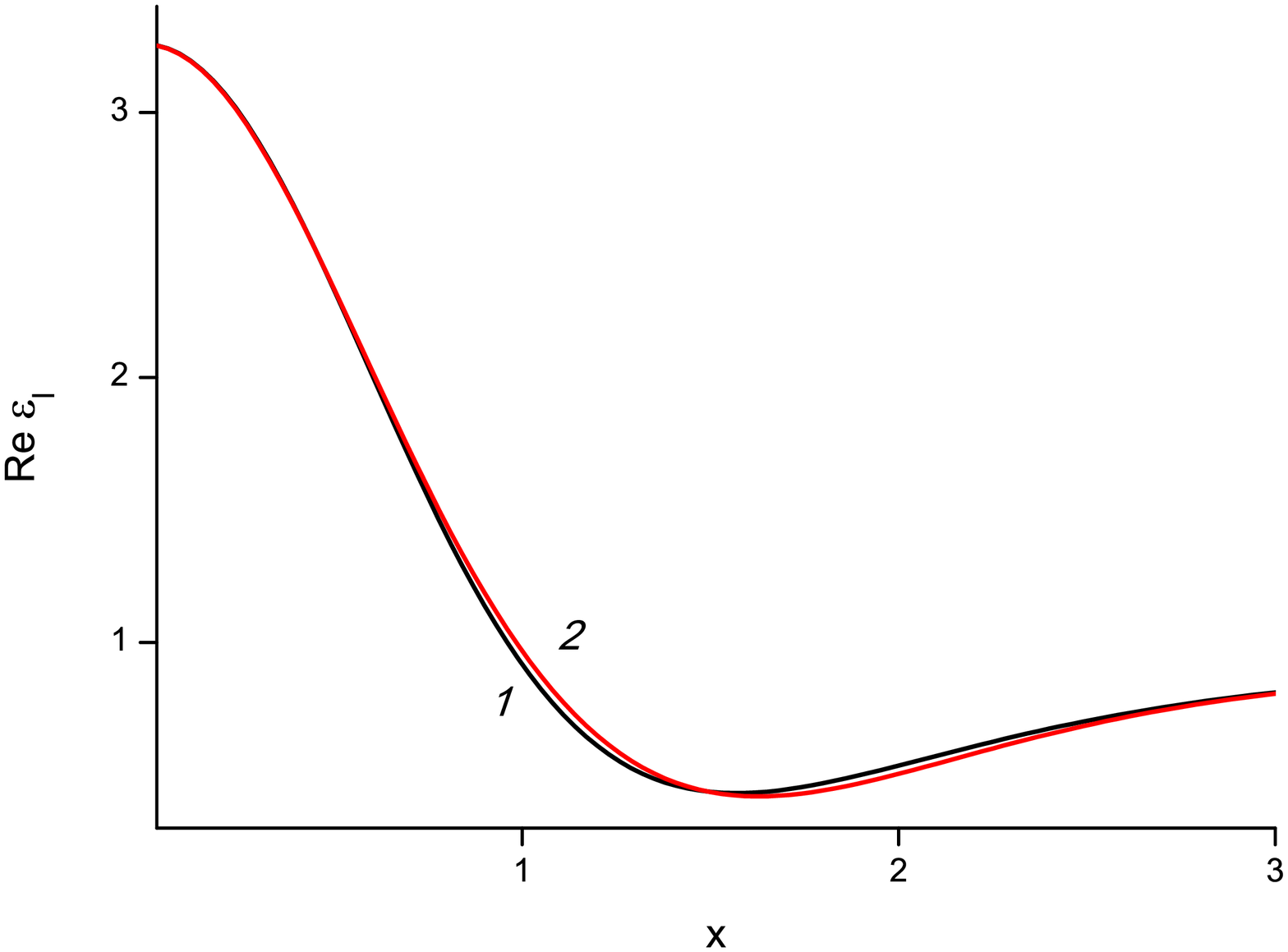}
\center{Fig. 7. Real part of dielectric function,
$Q=1, y=0.01$.}
\includegraphics[width=17.0cm, height=10cm]{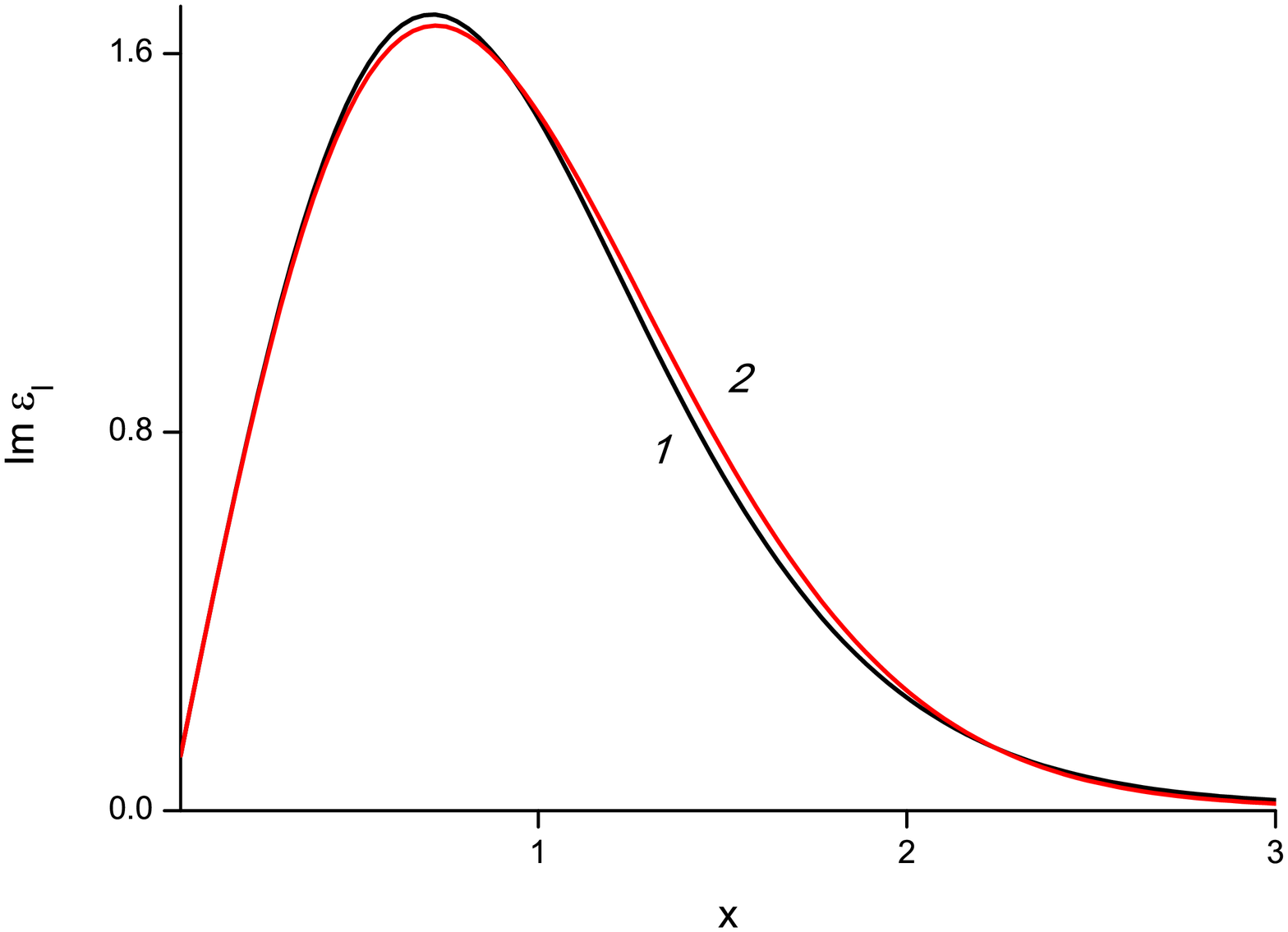}
\center{Fig. 8. Imaginary part of dielectric function,
$Q=1, y=0.01$.}
\end{figure}

\begin{figure}[t]\center
\includegraphics[width=16.0cm, height=9cm]{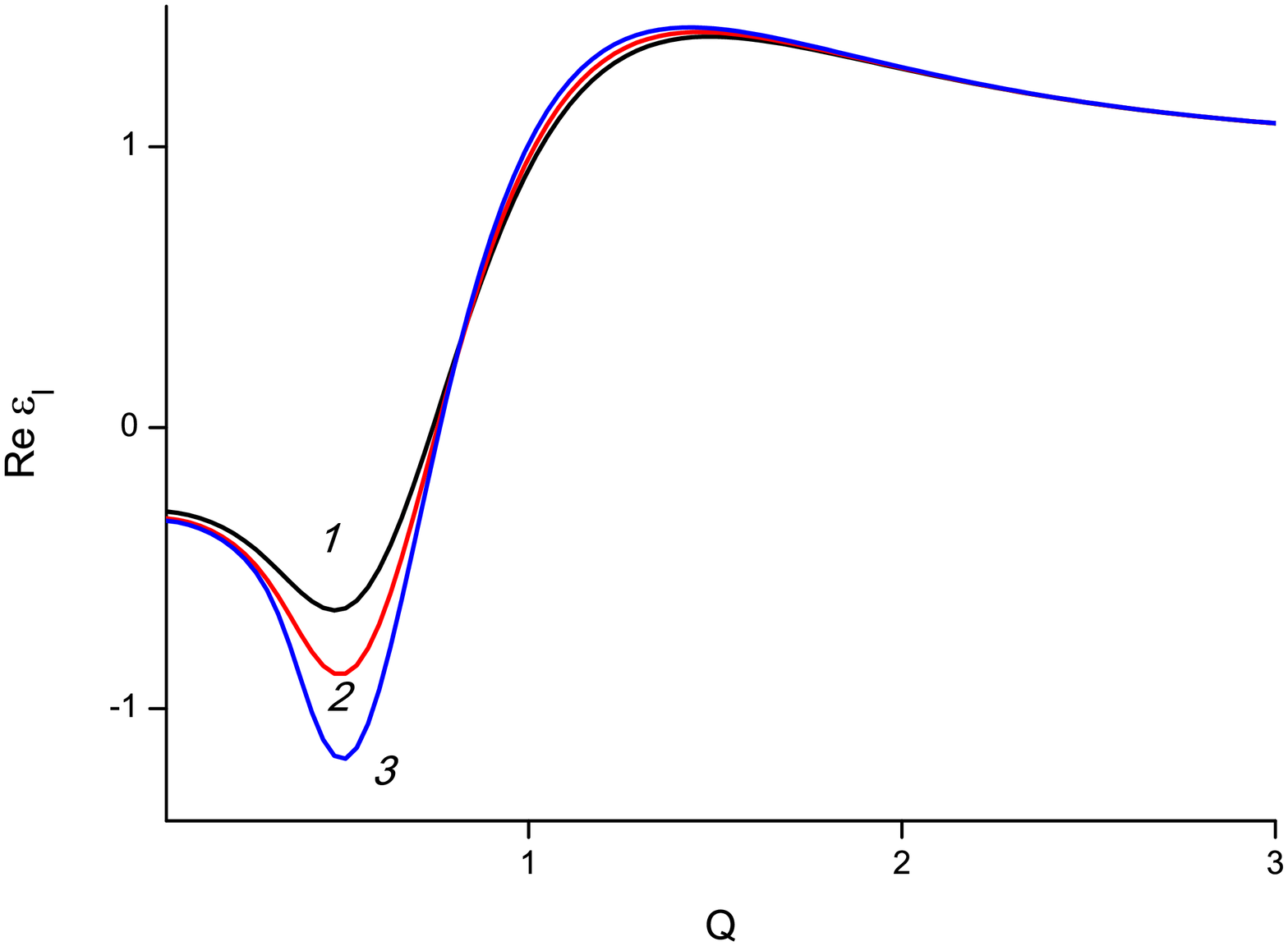}
\center{Fig. 9. Real part of dielectric function,
$x=1$. Maxwellian plasma.}
\includegraphics[width=17.0cm, height=9cm]{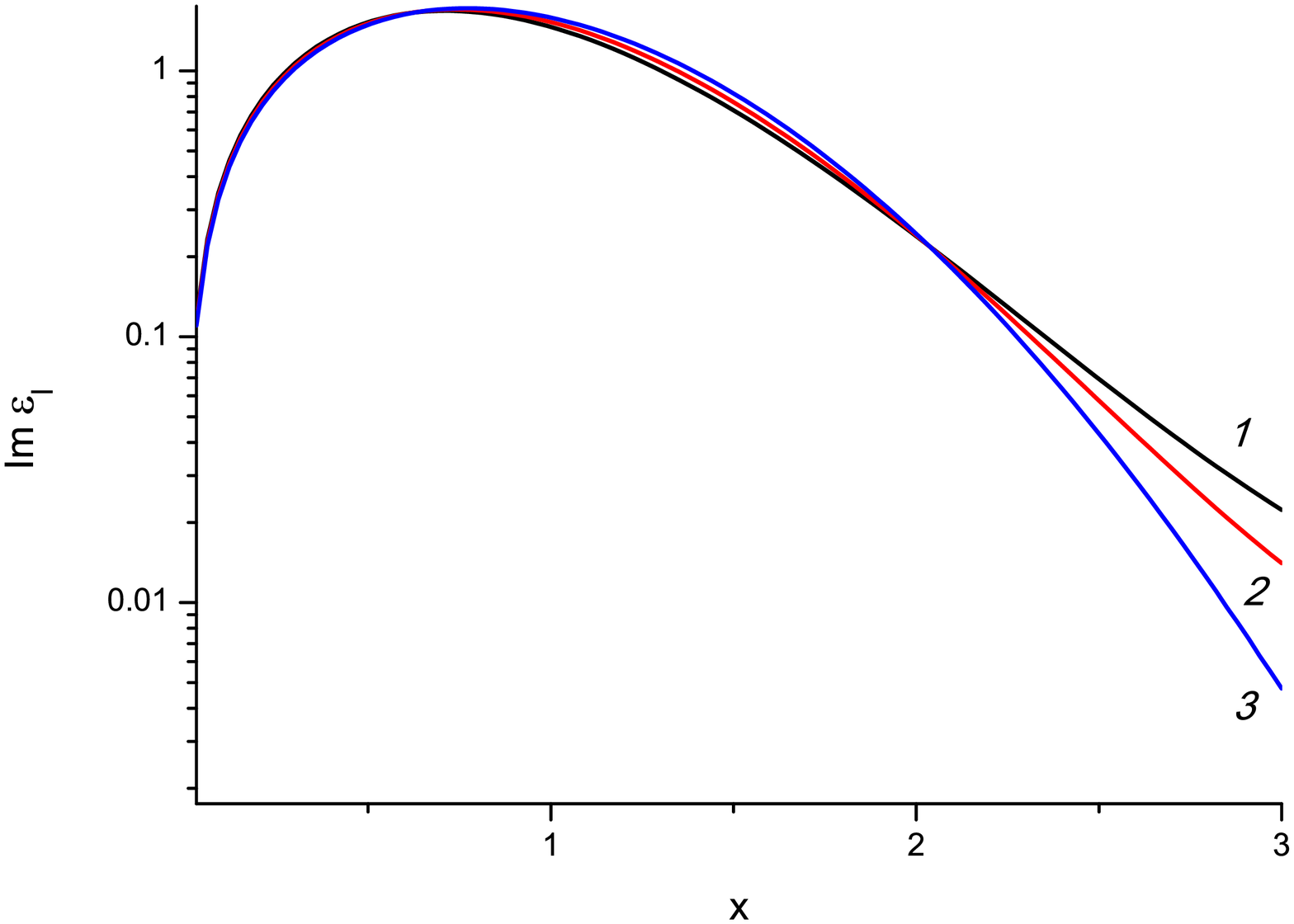}
\center{Fig. 10. Imaginary
 part of dielectric function,
$Q=1$.  Maxwellian plasma.}
\end{figure}

\begin{figure}[h]\center
\includegraphics[width=16.0cm, height=9cm]{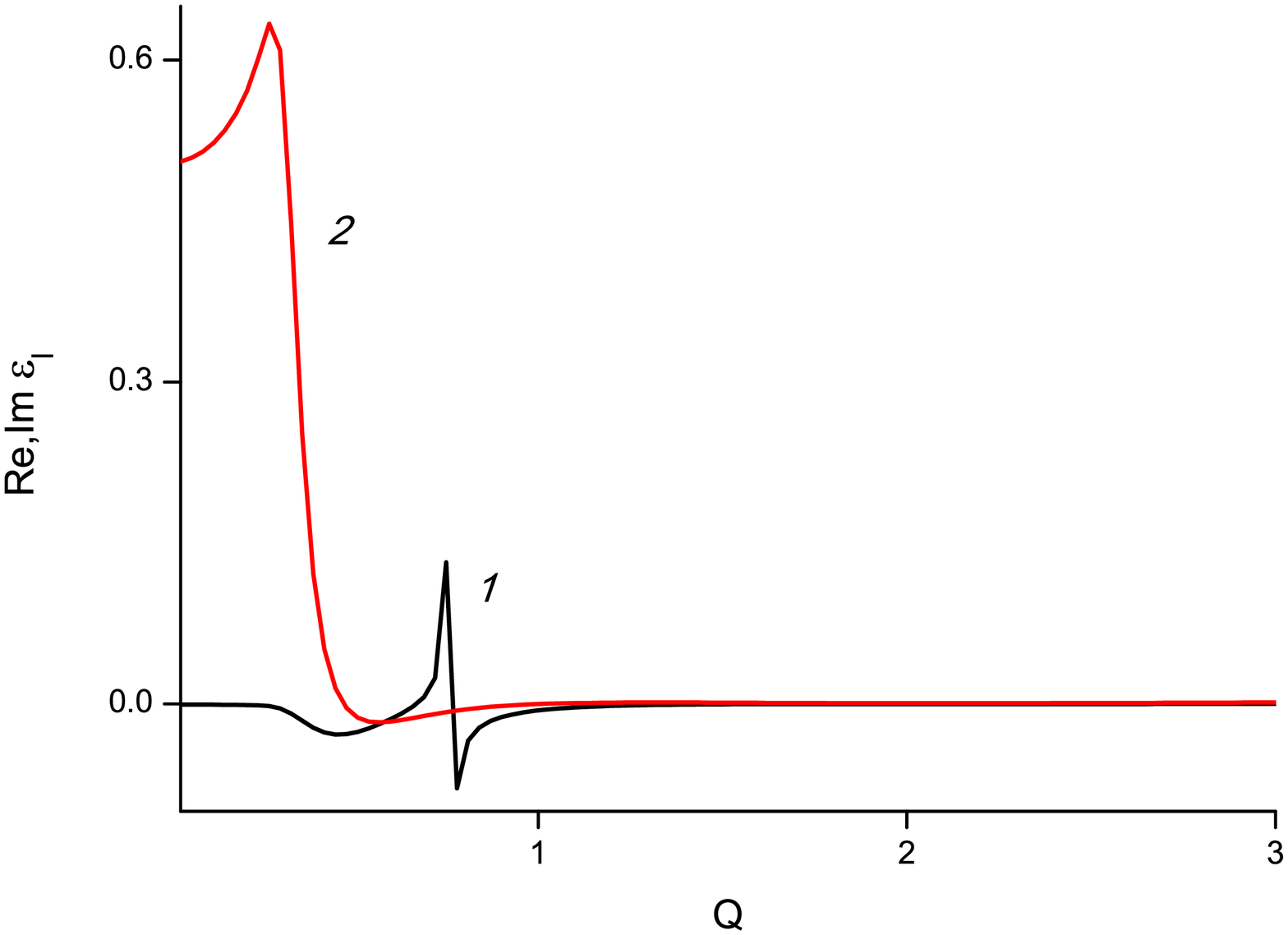}
\center{Fig. 11. Relative deviation of the real and imaginary parts
of dielectric function. Graphics $O_r(Q,1,0.01)$ (curve 1) и $O_i(Q,1,0.01)$\\
(curve 2).  Maxwellian plasma.}
\includegraphics[width=17.0cm, height=9cm]{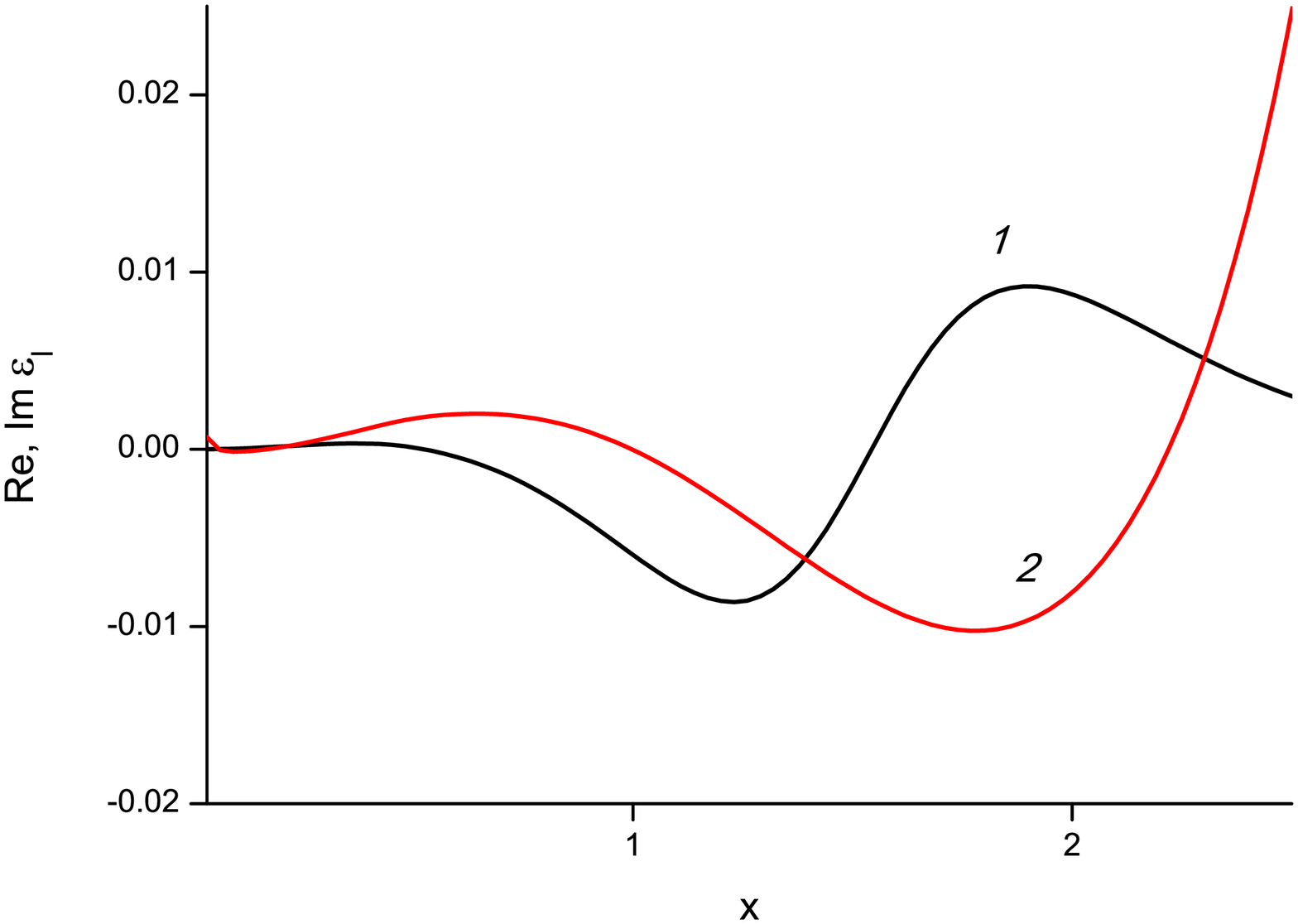}
\center{Fig. 12. Relative deviation of the real and imaginary parts
of dielectric function. Graphics $O_r(1,x,0.01)$ (curve 1) and $O_i(1,x,0.01)$\\
(curve 2).  Maxwellian plasma.}
\end{figure}

\end{document}